\documentclass[twocolumn,showpacs,superscriptaddress,preprintnumbers,amsmath,amssymb]{revtex4}
\usepackage{graphicx}% Include figure files
\usepackage{dcolumn}% Align table columns on decimal point
\usepackage{bm}% bold math

\begin{document}

%\preprint{APS/123-QED}

\title{Observation of three-dimensional bulk Fermi surfaces for a strongly correlated material by soft x-ray $h\nu$-dependent (700-860 eV) ARPES}% Force line breaks with \\

\author{M. Yano}
 %\altaffiliation[Also at ]{Division of Materials Physics, Graduate School of Engineering  Science, Osaka University, Toyonaka, Osaka 560-8531, Japan}%Lines break automatically or can be forced with \\
\author{A. Sekiyama}
\author{H. Fujiwara}
\author{T. Saita}
\author{S. Imada}
\affiliation{Division of Materials Physics, Graduate School of Engineering  Science, Osaka University, Toyonaka, Osaka 560-8531, Japan}

\author{T. Muro}
\affiliation{Japan Synchrotoron Radiation Research Institute, SPring-8, Mikazuki, Hyogo 679-5198, Japan}

\author{Y. Onuki}
\affiliation{Department of Physics, Graduate School of Science, Osaka University, Toyonaka, Osaka 560-0043, Japan}

\author{S. Suga}%
\affiliation{Division of Materials Physics, Graduate School of Engineering  Science, Osaka University, Toyonaka, Osaka 560-8531, Japan}

\date{\today}% It is always \today, today,

\begin{abstract}
Three-dimensional Fermi surfaces at a high temperature have been clarified for a strongly correlated Ce compound, ferromagnet CeRu$_2$Ge$_2$ in the paramagnetic phase, by virtue of a soft x-ray $h\nu$-dependent (700-860 eV) ARPES. Although the observed Fermi surfaces as well as quasiparticle
dispersions are partly explained by a band-structure calculation based on a localized $4f$ model, qualitative discrepancy in experiments, between our
APRES in the paramagnetic phase and de Haas-van Alphen measurement in the ferromagnetic phase, is revealed. This suggests a fundamental change in the
$4f$ contribution to the Fermi surfaces across the magnetic phase transition widely seen for Ce compounds.
\end{abstract}

\maketitle

Many macroscopic properties of solids such as resistivity, specific heat and susceptibility depend strongly on momentum distribution of electrons on
the chemical potential, namely, shapes and characters of the Fermi surfaces.
Therefore, detection of the Fermi surfaces is important to clarify the electronic properties of solids. The quantum oscillation measurement using de Haas-van Alphen (dHvA) effect is known as a powerful technique to observe the cross-sections of the Fermi surfaces\cite{AandM}.
The dHvA measurement has so far been applied also for many strongly correlated rare-earth materials\cite{CeNi, CeCu6, CeRuSi_2}.
Since the successful consistency between the experimentally observed Fermi surfaces and the band-structure calculation assuming itinerant $4f$ electrons for CeSn$_3$\cite{CeSn3_1, CeSn3LDA}, the dHvA mesurement has been recognized as a convincing tool to qualitatively judge whether the $4f$ electrons are "itinerant" or "localized" for the ground state of strongly correlated Ce compounds.
However, the Fermi surfaces at high temperature above several tens K, at which the electronic structures are often deviated from the ground state due to magnetic phase transitions and/or the Kondo effect, have not been experimentally clarified for many Ce compounds because a low temperature is required for the dHvA measurement.
The low-$h\nu$ angle-resolved photoemission spectroscopy (ARPES) is also useful to reveal the characters of the two-dimensional and/or surface Fermi surfaces as seen in the results for high-$T_C$ cuprates\cite{DamRMP}.
As for the rare-earth compounds, ARPES measurements for XRu$_2$Si$_2$ (X = La, Ce, Th, U) have been performed by using $h\nu$ in a 14-230 eV range\cite{fs_La-Ce}.
Recently high-energy ($h\nu > $500 eV) photoemission is found to be very effective in probing bulk states\cite{bulk, Si_Ge_res, High-energyARPES, Suga_B}.
In this letter, we demonstrate the power of soft x-ray $h\nu$-dependent ARPES for clarifing the bulk three-dimensional Fermi surface topology of a strongly correlated rare-earth compound, whose $4f$ electronic states are mutually different between the bulk and surface\cite{Si_Ge_res}.

\begin{figure}[htbp]
  \begin{center}
    \includegraphics[keepaspectratio=true,height=100mm]{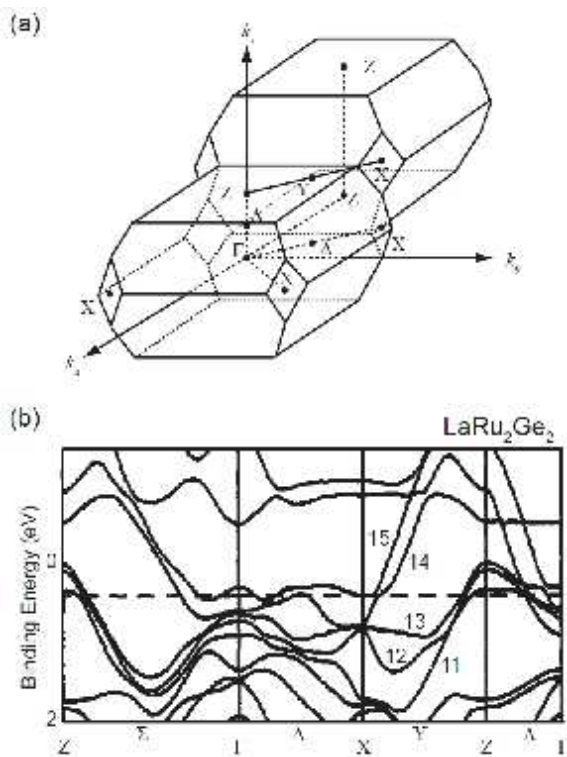}
  \end{center}
  \caption{(a)The Brillouin zone of the body-centered tetragonal crystal CeRu$_2$Ge$_2$ with $|\Gamma-Z|=2\pi /c$. (b) The band calculation with APW method for LaRu$_2$Ge$_2$\cite{Ge}. The Fermi surfaces are formed by 5 bands from 11 to 15.}%{}内にタイトルを記入してください
  \label{BZandCalc.eps}
\end{figure}

We have performed the ARPES measurements for CeRu$_2$Ge$_2$ which has one $4f$ electron per unit cell then we have compared with the band calculation for $4f$ electron localized model LaRu$_2$Ge$_2$.
CeRu$_2$Ge$_2$ is a requisite material to understand the electronic states of Ce heavy fermion systems because the rather localized $4f$ electrons become itinerant under high pressures\cite{HP}.
CeRu$_2$Ge$_2$ is a ferromagnet with $T_C\sim $8 K\cite{ele_heat, TC} and the $4f$ electrons of CeRu$_2$Ge$_2$ are thought to be localized because RKKY interaction is dominant compared with the Kondo effect at low temperatures with the electronic specific heat coefficient of about 20 mJ/mol K$^2$\cite{ele_heat} and the Kondo temperature $T_K <$ 1 K\cite{neutron, KondoTemp}.
CeRu$_2$Ge$_2$ crystallizes in a tetragonal ThCr$_2$Si$_2$-type structure with a=4.268 \AA \ and c=10.07 \AA \ at 18 K\cite{ele_heat} and its Brillouin zone is shown in Fig. \ref{BZandCalc.eps}(a).
On the other hand, isostructural CeRu$_2$Si$_2$ has itinerant $4f$ electrons with its low-temperature electronic specific heat coefficient of about 350 mJ/mol K$^2$ \cite{Si_ele_heat} and $T_K \sim $20 K\cite{KondoTemp}. 
These properties are consistent with the bulk sensitive $3d$ - $4f$ resonant photoemission results\cite{Si_Ge_res}.
The larger lattice constant of CeRu$_2$Ge$_2$ than that of CeRu$_2$Si$_2$ makes such different $4f$ electron properties.

CeRu$_2$Ge$_2$ single crystal was grown by the Czochralski pulling method.
The high energy ARPES measurements have been performed from $h\nu $=700 to 860 eV with an energy step of 5 eV at BL25SU in SPring-8\cite{SR}.
The light incidence angle was $\sim 45$ degrees with respect to the sample surface normal.
The base pressure was about $3\times 10^{-8}$ Pa.
We have performed the measurements at 20 K, where the sample was in the paramagnetic phase.
The clean surface was obtained by cleaving \textit{in situ} providing a (001) plane.
A GAMMADATA-SCIENTA SES200 analyzer was used covering more than a whole Brillouin zone along the  direction of the slit.
The energy resolution was set to $\sim $200 meV for Fermi surface mapping.
The cleanliness was confirmed by the absence of the O 1\textit{s} and C 1\textit{s} photoemission signals.
First, we have performed $k_z-k_{xy}$ mapping at several $h\nu$ and angles.
After determining the $h\nu $ corresponding to the high symmetry points along the [001] direction, we have performed detailed angle-dependent ARPES for $k_x-k_y$ mapping.
Then we have performed $h\nu$ dependent ARPES for $k_z-k_{xy}$ mapping through the high symmetry points in the $k_{xy}$ Brillouin zone.

In order to experimentally determine the exact value of $|k|$, we have taken the incident photon momentum into account.
If x-ray was incident onto a sample at 45 degrees with respect to the surface normal, for example, this incident photon has the momentum parallel ($q_{\|  }$) and perpendicular ($q_{\bot }$) to the surface, which are $2\pi /\lambda /\sqrt{2}$ (\AA $^{-1}$)=$2\pi h\nu $(eV)$/ 12398 /\sqrt{2}$.
In the case of $h\nu = 700$ eV, the photon momentum values of both $| q_{\|} |$ and $ |q_{\bot }| $ are about 0.25 \AA$^{-1}$.
Because the value $|k_z|$ of the 1st Brillouin zone of CeRu$_2$Ge$_2$ ($2\pi /c \sim 0.62$ \AA$^{-1}$) is comparable to the value of the photon momentum, both $q_{\|}$ and $q_{\bot }$ cannot be negligible.

% 図の挿入

%バンド計算とバンド15の組成
The observed ARPES data have been compared with the band calculation of paramagnetic LaRu$_2$Ge$_2$ performed by H. Yamagami and A. Hasegawa\cite{Ge} with using a symmetrized relativistic augmented-plane wave (APW) method\cite{APW}.
According to their calculation, five bands (11 - 15) cut the Fermi level ($E_F$) (Fig. \ref{BZandCalc.eps}(b)) forming five Fermi surfaces which are composed of the La $4d$ and Ru $4d$ states.
Five Fermi surfaces are likewise derived for the itinerant $4f$ electron system CeRu$_2$Si$_2$\cite{Ce}.
Between the localized LaRu$_2$Ge$_2$ and itinerant CeRu$_2$Si$_2$, the shapes of Fermi surfaces derived from the bands 11 to 14 are not much different but the shape of the band 15 is mutually different.
Among the calculated bands of LaRu$_2$Ge$_2$, the lower four bands from 11 to 14 form the hole-like Fermi surfaces which are centered at the Z point in the Brillouin zone, and the highest band 15 forms the electron-like Fermi sheet.

\begin{figure}
     \includegraphics[keepaspectratio=true,height=100mm]{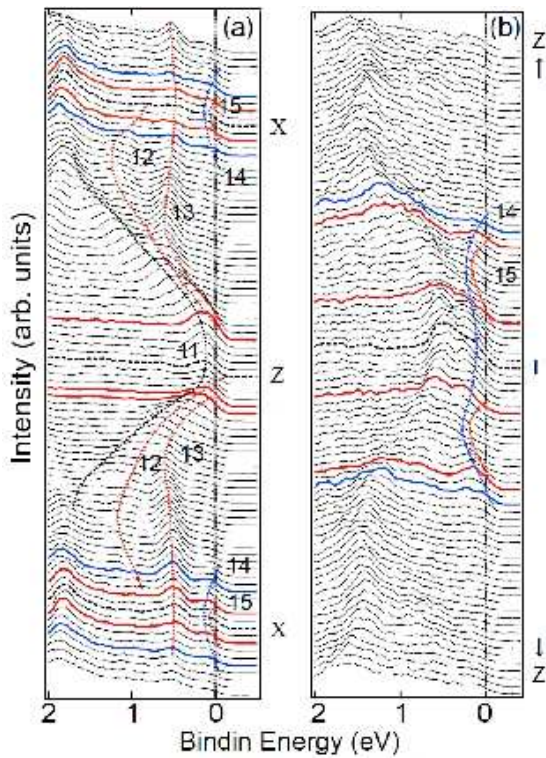}
  \caption{(color online) ARPES spectra near $E_F$ of CeRu$_2$Ge$_2$ with an energy resolution of $\sim$ 200 meV. The dashed lines represent the bands corresponding to the APW calculation for LaRu$_2$Ge$_2$. (a) EDCs along the X - Z - X cut ($h\nu $=755 eV) shown in the third zone from the left in Fig. \ref{BZandCalc.eps}(b). (b) EDCs along the Z - $\Gamma $ - Z cut ($h\nu $=820 eV) shown in the leftmost zone of Fig. \ref{BZandCalc.eps}(b).}%{}内にタイトルを記入してください
  \label{edcs}
\end{figure}

%EDCの説明
Figure \ref{edcs}(a) shows the energy distribution curves (EDCs) along the in-plane X - Z - X direction for CeRu$_2$Ge$_2$.
The positions of the wave number where each band cuts $E_F$ is also estimated by the momentum distribution curves(MDCs).
Five bands corresponding to the bands 11 to 15 predicted by the band calculation for LaRu$_2$Ge$_2$ are clearly seen.
There is a clear peak near $E_F$ at the Z point in Fig. \ref{edcs}(a).
The presence of this peak is in a strong contrast to the result of the band-structure calculation for LaRu$_2$Ge$_2$, which predicts no quasiparticle peak near $E_F$ at the Z point in Fig. 1(b).
It is thus revealed that the band 11 is located on the occupied side and does not form the Fermi surface.
Although the band 15 does not cut $E_F$ at the X point in the calculation for LaRu$_2$Ge$_2$, our experimental results show that the band 15 exists on the occupied side at the X point.
Figure \ref{edcs}(b) shows EDCs along the in-plane Z - $\Gamma$ - Z direction.
The comparison between Fig. \ref{BZandCalc.eps} (b) and Fig. \ref{edcs} has shown the similarity of the observed bands 15 and 14 in CeRu$_2$Ge$_2$ with the result of the band calculation of LaRu$_2$Ge$_2$.

\begin{figure}
\includegraphics[keepaspectratio=true,height=70mm]{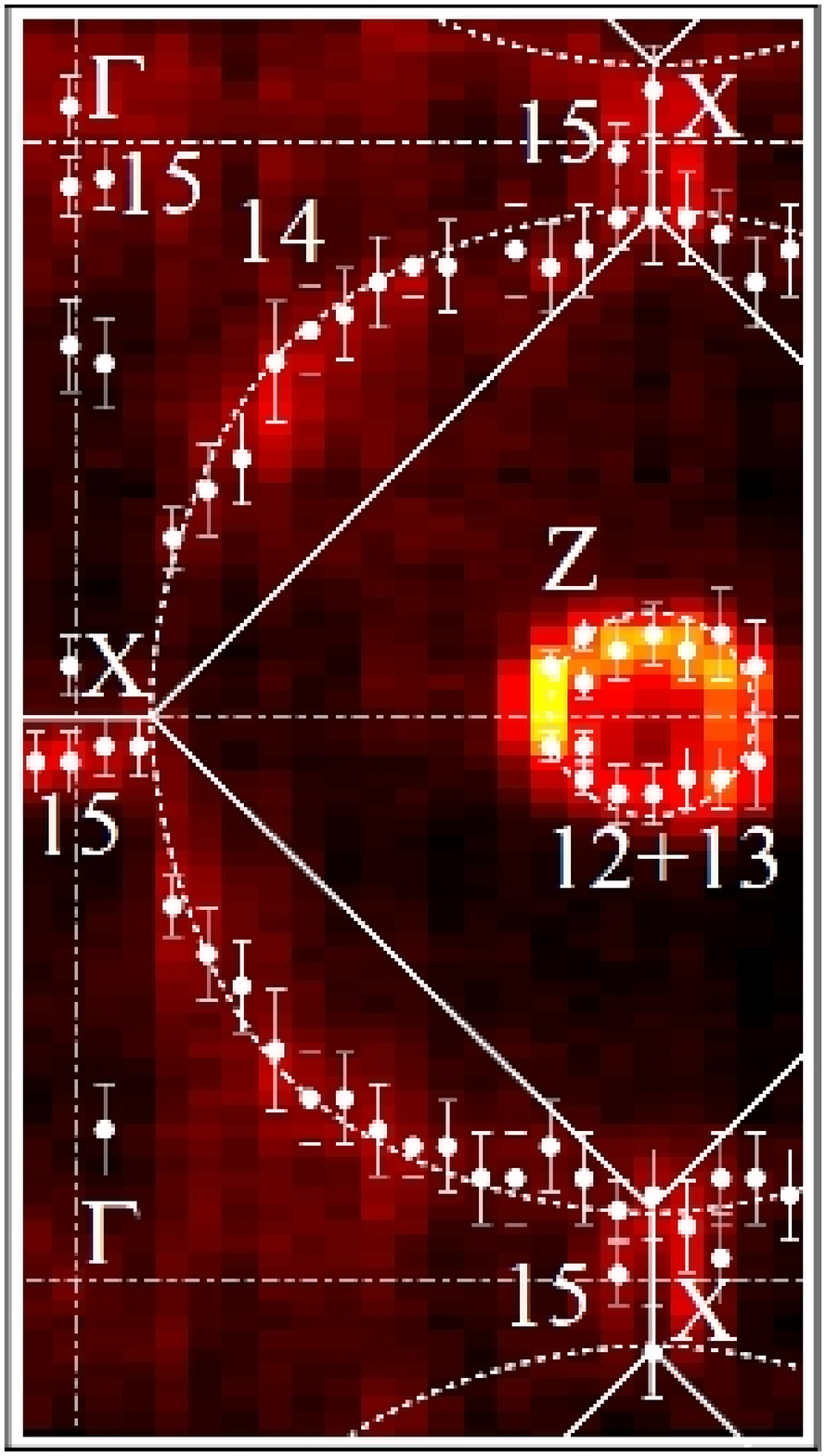}% Here is how to import EPS art
\caption{(color online) Fermi surface slice of CeRu$_2$Ge$_2$ at $k_z \sim 2\pi /c$ ($h\nu $= 755 eV). Photoemission intensity map was obtained by integrating the PES intensity from +0.1 eV to -0.1 eV. The solid lines represent the corresponding Brillouin zone and the dashed lines represent high symmetry lines. White dots (with error bars) represent the estimated $k_F$ from EDCs and MDCs.}
\label{fig:fs755line.eps}
\end{figure}

%マッピング755の説明
Figure \ref{fig:fs755line.eps} displays Fermi surfaces' topology at $k_z \sim 2\pi /c$ by integrating the intensity around $E_F$ of EDCs.
The X - Z - X line corresponds to Fig. \ref{edcs}(a).
The intensities around the Z point are due to the bands 11-13.
We have clearly observed the small Fermi surface contours around the Z point derived from both bands 12 and 13 and the large Fermi surface contour derived from the band 14 as judged from Fig. 2(a).
Furthermore, we recognize a small Fermi surface contour centered at the X point.
The possible origin of this small Fermi surface is the band 15, whereas its $E_F$ crossing is not predicted by the calculation for LaRu$_2$Ge$_2$.

\begin{figure}
    \includegraphics[keepaspectratio=true,height=80mm]{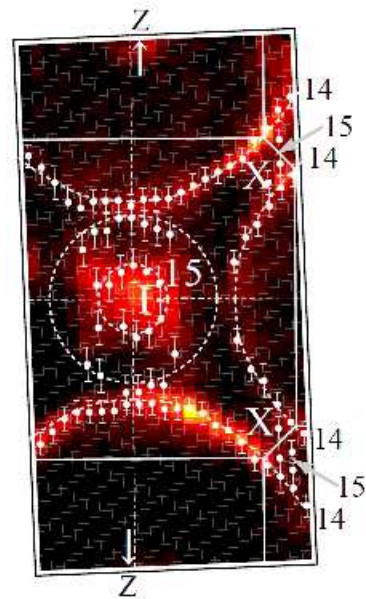}
  \caption{(color online) Fermi surface slice of CeRu$_2$Ge$_2$ at $k_z \sim 0$ ($h\nu $= 820 eV). Photoemission intensity map was obtained by integrating the PES intensity from +0.1 eV to -0.1 eV.}%{}内にタイトルを記入してください
  \label{fig:FS820.eps}
\end{figure}

%マッピング820の説明
Figure \ref{fig:FS820.eps} demonstrates Fermi surfaces' topology at $k_z \sim 0$.
The Z - $\Gamma$ - Z line corresponds to Fig. \ref{edcs}(b).
The very large Fermi surface derived from the band 14 is clearly seen around the Z points.
Furthermore a part of the Fermi surface derived from the band 15, whose shape looks like a doughnut centered at the $\Gamma $ point according to the calculation for LaRu$_2$Ge$_2$, can be traced.

\begin{figure}
     \includegraphics[keepaspectratio=true,height=39mm]{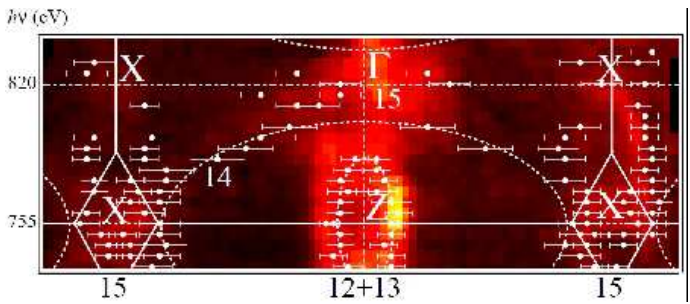}
  \caption{(color online) Fermi surface slice in $k_z$(ordinate) - $k_{xy}$(abscissa) plane. Photon energies were changed from 735 eV to 840 eV. $h\nu =$820 eV and 755 eV correspond to the $\Gamma $ point and the Z points, respectively, along the $\Gamma $ - Z direction corresponding to the right most zone in Fig.\ref{BZandCalc.eps}(b). The upper horizontal X - $\Gamma$ -X axis corresponds to the second zone from  the left in Fig. \ref{BZandCalc.eps}(b). Photoemission intensity map was obtained by integrating the PES intensity from +0.1 eV to -0.1 eV.}%{}内にタイトルを記入してください
  \label{fig:FSkz.eps}
\end{figure}

%kz-kxyマッピングの説明
Meanwhile, Fig. \ref{fig:FSkz.eps} shows the Fermi surface slice at $k_z-k_{xy}$ plane measured with an energy step of 5 eV.
In this figure are recognized the Fermi surfaces 12 and 13 centered at the Z point, which are prolongated along the $k_z$ direction.
The very large Fermi surface 14 compressed vertically can be also experimentally traced.
Along the horizontal X - $\Gamma$ - X direction, the band 15 cuts $E_F$ near the $\Gamma$ and X points.
The Fermi surface of the band 15 is only partly observed near the $\Gamma$ point because of the noticeable background.
The nonnegligible intensities around the $\Gamma $ point and the Z point suggest the proximity of the band 14 and the band 11, respectively, which are not crossing $E_F$ near these points but staying in the vicinity of $E_F$ as we can see in Fig. \ref{edcs}.
We have also observed the Fermi surface continuing along the ordinate X - X axis derived from the band 15.
In our Fermi surface mapping, this Fermi surface along the X - X axis in Fig. \ref{fig:FSkz.eps} is also detected near the X point in Fig. \ref{fig:fs755line.eps} corresponding to the upper X - Y - Z plane (Fig. \ref{BZandCalc.eps}(a)) and also near the X point in Fig. \ref{fig:FS820.eps} corresponding to the middle $\Gamma$ - $\Delta$ - X plane (Fig. \ref{BZandCalc.eps}(a)).

%フェルミ面を総括して
The band calculation for LaRu$_2$Ge$_2$ predicts that there are three hole pockets derived from the band 11-13 and a large hole Fermi surface derived from the band 14 centered at the Z point.
In addition it predicts that the band 15 forms both a doughnut-like electron pocket centered at the $\Gamma$ point and a discontinuous Fermi surface along the ordinate X - X direction.
On the other hand, the dHvA measurements for CeRu$_2$Ge$_2$ in the ferromagnetic phase have confirmed the existence of all Fermi surface sheets spin-split\cite{dHvA, dHvA_1} corresponding to the bands 11-15 predicted for LaRu$_2$Ge$_2$.
The agreement between the experimental Fermi surfaces by dHvA and ARPES measurements and the calculated results is quite good for the bands 12, 13 and 14.

%%%%%%%考察%%%%%%%%%%%%%%%%%%%%%%%%%%%%%%%%%%%%
From our ARPES results, however, the Fermi surface of CeRu$_2$Ge$_2$ derived from the band 11 is found not to exist as revealed in Fig. \ref{edcs}(a).
It is also found that the band 15 has a continuous Fermi surface along the $k_z$ direction in the paramagnetic phase.
These results are in a strong contrast to the dHvA results in the ferromagnetic phase.
Although the Fermi surface of CeRu$_2$Ge$_2$ in the ferromagnetic phase is similar to that of LaRu$_2$Ge$_2$, the difference of our ARPES results from them are consistently understood if $E_F$ of CeRu$_2$Ge$_2$ in the paramagnetic phase is energetically higher than that of the calculation for LaRu$_2$Ge$_2$.
$E_F$ shift of CeRu$_2$Ge$_2$ in the paramagnetic phase from LaRu$_2$Ge$_2$ or CeRu$_2$Ge$_2$ in the ferromagnetic phase is thought to be due to the increased number of the electrons contributing to the near $E_F$ bands in CeRu$_2$Ge$_2$, where the weak but nonnegligible hybridization of the Ce $4f$ electron should be additionally taken into account in the paramagnetic phase.
The difference of the electric resistivity between CeRu$_2$Ge$_2$ and LaRu$_2$Ge$_2$ is suddenly diminished below $T_C$ of CeRu$_2$Ge$_2$\cite{ele_heat}, indicating the reduction of electron scattering by the ferromagnetic ordering.
This suggests that the contribution of the $4f$ electrons to the Fermi surfaces due to the hybridization is reduced in the ferromagnetic phase.
Then the number of the electrons contributing to the Fermi surfaces decreases below $T_C$.
Accordingly, the band 11 crosses $E_F$ near the Z point and the band 15 might form discontinuous Fermi surfaces along the X - X ($k_z$) direction as predicted by the band-structure calculation.

%%%%%%%%%%%%%結論まとめ%%%%%%%%%%%%%%%%%%%%%%%%%%%%%%
We have performed three-dimensional bulk-sensitive ARPES measurements for paramagnetic CeRu$_2$Ge$_2$ by using soft x-rays.
Although the Fermi surfaces obtained for the bands 12, 13 and 14 are in good agreement with the result of the band calculation for paramagnetic LaRu$_2$Ge$_2$, the predicted band 11 is found to be not contributing to the Fermi surface in paramagnetic CeRu$_2$Ge$_2$.
The band 15 is confirmed to have the doughnut-like shape around the $\Gamma$ point whereas a rod-like continuous Fermi surface along the X - X axis is observed in a strong contrast to the band calculation.
The slight hybridization in the paramagnetic phase and the magnetic ordering below $T_C$ are thought to be essential to understand the behaviors of three-dimensional electronic structures of CeRu$_2$Ge$_2$.
 
%%%%Acknowledgment%%%%%%%%%%%%%%%%%%%%%%%%
We are grateful to H. Yamagami for fruitful discussions.
We thank T. Miyamachi and H. Higahsimichi for their help in the experiments.
The present work was performed at SPring-8 under the proposal (2004A6009-NS-np, 2004B0400-NSa-np) supported by the Grant-in-Aids for Creative Scientific Research (15GS0213) of MEXT Japan and 21st Century COE program (G18) of Japan Society for Promotion of Science.

\end{document}